\documentclass[twocolumn,showpacs,preprintnumbers,superscriptaddress,amsmath,amssymb,PRL]{revtex4-2}
\usepackage{graphicx}
\usepackage{dcolumn}
\usepackage{float}
\usepackage{mathptmx, courier, pifont}
\usepackage[scaled=0.92]{helvet}
\usepackage[T1]{fontenc}
\usepackage{textcomp}
\usepackage{color}
\usepackage[normalem]{ulem}
\usepackage{algorithmic} 

\usepackage[dvipsnames]{xcolor}
\usepackage[colorlinks=true,urlcolor=Blue,linkcolor=Blue,citecolor=blue]{hyperref}
\usepackage[all]{hypcap}
\usepackage{makecell}
\usepackage{footmisc}
\usepackage{bibunits}
\defaultbibliographystyle{apsrev4-1_titled}
\defaultbibliography{my-final-bib-from-jabref_titles}
\usepackage{xcolor}
\definecolor{Mycolor}{HTML}{bfffbf}

\begin{document}
\begin{bibunit}


%
%

\title{High-precision mass measurements of neutron deficient silver isotopes \\ probe the robustness of the $N$ = 50 shell closure}

\author{Zhuang~Ge}\thanks{Corresponding author: zhuang.z.ge@jyu.fi}
\affiliation{Department of Physics, University of Jyv\"askyl\"a, P.O. Box 35, FI-40014, Jyv\"askyl\"a, Finland}
\affiliation{GSI Helmholtzzentrum f\"ur Schwerionenforschung GmbH, 64291 Darmstadt, Germany}
\author{Mikael~Reponen}\thanks{Corresponding author: mikael.h.t.reponen@jyu.fi}
\affiliation{Department of Physics, University of Jyv\"askyl\"a, P.O. Box 35, FI-40014, Jyv\"askyl\"a, Finland}%
\author{Tommi~Eronen}
\affiliation{Department of Physics, University of Jyv\"askyl\"a, P.O. Box 35, FI-40014, Jyv\"askyl\"a, Finland}%
\author{Baishan~Hu}
\affiliation{TRIUMF, 4004 Wesbrook Mall, Vancouver, BC V6T 2A3, Canada}
\affiliation{National Center for Computational Sciences, Oak Ridge National Laboratory, Oak Ridge, Tennessee 37831, USA}
\affiliation{Physics Division, Oak Ridge National Laboratory, Oak Ridge, Tennessee 37831, USA}
\author{Markus Kortelainen}
\affiliation{Department of Physics, University of Jyv\"askyl\"a, P.O. Box 35, FI-40014, Jyv\"askyl\"a, Finland}
\author{Anu~Kankainen}
\affiliation{Department of Physics, University of Jyv\"askyl\"a, P.O. Box 35, FI-40014, Jyv\"askyl\"a, Finland}%
\author{Iain~Moore}
\affiliation{Department of Physics, University of Jyv\"askyl\"a, P.O. Box 35, FI-40014, Jyv\"askyl\"a, Finland}%
\author{Dmitrii~Nesterenko}
\affiliation{Department of Physics, University of Jyv\"askyl\"a, P.O. Box 35, FI-40014, Jyv\"askyl\"a, Finland}%
\author{Cenxi~Yuan}
\affiliation{Sino-French Institute of Nuclear Engineering and Technology, Sun Yat-Sen University, Zhuhai, 519082, Guangdong, China}
\author{Olga~Beliuskina}
\affiliation{Department of Physics, University of Jyv\"askyl\"a, P.O. Box 35, FI-40014, Jyv\"askyl\"a, Finland}%
\author{Laetitia~Ca{\~n}ete}
\altaffiliation{Northeastern University London, Devon House, 58 St Katharine's Way, E1W 1LP, London, United Kingdom}
\affiliation{Department of Physics, University of Jyv\"askyl\"a, P.O. Box 35, FI-40014, Jyv\"askyl\"a, Finland}%
\author{Ruben~de~Groote}
\affiliation{Department of Physics, University of Jyv\"askyl\"a, P.O. Box 35, FI-40014, Jyv\"askyl\"a, Finland}%
\affiliation{KU Leuven, Instituut voor Kern- en Stralingsfysica, B-3001 Leuven, Belgium}
\author{Celement~Delafosse}
\affiliation{Department of Physics, University of Jyv\"askyl\"a, P.O. Box 35, FI-40014, Jyv\"askyl\"a, Finland}%
\affiliation{Universit\'e Paris-Saclay, CNRS/IN2P3, IJCLab, 91405 Orsay, France}
\author{Pierre~Delahaye}
\affiliation{GANIL, CEA/DSM-CNRS/IN2P3, Bd Henri Becquerel, 14000 Caen, France}%
\author{Timo~Dickel}
\affiliation{GSI Helmholtzzentrum f\"ur Schwerionenforschung GmbH, 64291 Darmstadt, Germany}
\affiliation{II. Physikalisches Institut, Justus-Liebig-Universit\"at Gie\ss en, 35392 Gie\ss en, Germany}
\author{Antoine~de Roubin}
\altaffiliation{LPC Caen, Université de Caen Normandie, 14050 Caen Cedex, France}
\affiliation{Department of Physics, University of Jyv\"askyl\"a, P.O. Box 35, FI-40014, Jyv\"askyl\"a, Finland}%
\author{Sarina~Geldhof}
\affiliation{Department of Physics, University of Jyv\"askyl\"a, P.O. Box 35, FI-40014, Jyv\"askyl\"a, Finland}%
\affiliation{GANIL, CEA/DSM-CNRS/IN2P3, Bd Henri Becquerel, 14000 Caen, France}%
\author{Wouter~Gins}
\affiliation{Department of Physics, University of Jyv\"askyl\"a, P.O. Box 35, FI-40014, Jyv\"askyl\"a, Finland}%
\author{Jason~Holt}
\affiliation{TRIUMF, 4004 Wesbrook Mall, Vancouver, BC V6T 2A3, Canada}
\affiliation{Department of Physics, McGill University, Montr\'eal, QC H3A 2T8, Canada}
\author{Marjut~Hukkanen}
\affiliation{Department of Physics, University of Jyv\"askyl\"a, P.O. Box 35, FI-40014, Jyv\"askyl\"a, Finland}%
\affiliation{Universit\'e de Bordeaux, CNRS/IN2P3, LP2I Bordeaux, UMR 5797, F-33170 Gradignan, France}
\author{Arthur~Jaries}
\affiliation{Department of Physics, University of Jyv\"askyl\"a, P.O. Box 35, FI-40014, Jyv\"askyl\"a, Finland}%
\author{Ari~Jokinen} 
\affiliation{Department of Physics, University of Jyv\"askyl\"a, P.O. Box 35, FI-40014, Jyv\"askyl\"a, Finland}%
\author{\'Agota~Koszor\'us}
\affiliation{KU Leuven, Instituut voor Kern- en Stralingsfysica, B-3001 Leuven, Belgium}
\affiliation{Department of Physics, University of Liverpool, Liverpool, L69 7ZE, United Kingdom}
\affiliation{Experimental Physics Department, CERN, CH-1211 Geneva 23, Switzerland}
\author{Gabriella~Kripk\'o-Koncz}
\affiliation{II. Physikalisches Institut, Justus-Liebig-Universit\"at Gie\ss en, 35392 Gie\ss en, Germany}
\author{Sonja~Kujanp\"a\"a}
\affiliation{Department of Physics, University of Jyv\"askyl\"a, P.O. Box 35, FI-40014, Jyv\"askyl\"a, Finland}%
\author{Yihua~Lam}
\affiliation{Institute of Modern Physics, Chinese Academy of Sciences, Lanzhou 730000, People’s Republic of China}
\affiliation{School of Nuclear Science and Technology, University of Chinese Academy of Sciences, Beijing 100049, People’s Republic of China}
\author{Stylianos~Nikas}
\affiliation{Department of Physics, University of Jyv\"askyl\"a, P.O. Box 35, FI-40014, Jyv\"askyl\"a, Finland}%
\author{Alejandro~Ortiz-Cortes}
\affiliation{Department of Physics, University of Jyv\"askyl\"a, P.O. Box 35, FI-40014, Jyv\"askyl\"a, Finland}%
\affiliation{GANIL, CEA/DSM-CNRS/IN2P3, Bd Henri Becquerel, 14000 Caen, France}%
\author{Heikki~Penttil\"a}
\affiliation{Department of Physics, University of Jyv\"askyl\"a, P.O. Box 35, FI-40014, Jyv\"askyl\"a, Finland}%
\author{Daniel~Pitman-Weymouth}
\affiliation{Department of Physics, University of Jyv\"askyl\"a, P.O. Box 35, FI-40014, Jyv\"askyl\"a, Finland}%
\author{Wolfgang~Pla\ss}
\affiliation{GSI Helmholtzzentrum f\"ur Schwerionenforschung GmbH, 64291 Darmstadt, Germany}
\affiliation{II. Physikalisches Institut, Justus-Liebig-Universit\"at Gie\ss en, 35392 Gie\ss en, Germany}
\author{Ilkka~Pohjalainen}
\affiliation{Department of Physics, University of Jyv\"askyl\"a, P.O. Box 35, FI-40014, Jyv\"askyl\"a, Finland}%
\author{Andrea~Raggio}
\affiliation{Department of Physics, University of Jyv\"askyl\"a, P.O. Box 35, FI-40014, Jyv\"askyl\"a, Finland}%
\author{Sami~Rinta-Antila}
\affiliation{Department of Physics, University of Jyv\"askyl\"a, P.O. Box 35, FI-40014, Jyv\"askyl\"a, Finland}%
\author{Jorge~Romero}
\affiliation{Department of Physics, University of Jyv\"askyl\"a, P.O. Box 35, FI-40014, Jyv\"askyl\"a, Finland}%
\affiliation{Department of Physics, University of Liverpool, Liverpool, L69 7ZE, United Kingdom}
\author{Marek~Stryjczyk}
\affiliation{Department of Physics, University of Jyv\"askyl\"a, P.O. Box 35, FI-40014, Jyv\"askyl\"a, Finland}%
\author{Markus~Vilen}
\affiliation{Department of Physics, University of Jyv\"askyl\"a, P.O. Box 35, FI-40014, Jyv\"askyl\"a, Finland}%
\affiliation{Experimental Physics Department, CERN, CH-1211 Geneva 23, Switzerland}
\author{Ville~Virtanen}
\affiliation{Department of Physics, University of Jyv\"askyl\"a, P.O. Box 35, FI-40014, Jyv\"askyl\"a, Finland}%
\author{Alexandra~Zadvornaya}\thanks{Present address: University of Edinburgh, Edinburgh, EH9 3FD, United Kingdom}
\affiliation{Department of Physics, University of Jyv\"askyl\"a, P.O. Box 35, FI-40014, Jyv\"askyl\"a, Finland}%

\date{\today}

\begin{abstract}

High-precision mass measurements of exotic $^{95-97}$Ag isotopes close to the $N = Z$ line have been conducted with the JYFLTRAP double Penning trap mass spectrometer, with  the silver ions produced using the recently commissioned inductively-heated hot cavity catcher laser ion source at the Ion Guide Isotope Separator On-Line facility. 
The atomic mass of $^{95}$Ag was directly determined for the first time. In addition, the atomic masses of $\beta$-decaying 2$^+$ and 8$^+$ states in $^{96}$Ag have been identified and measured for the first time, and the precision of the $^{97}$Ag mass has been improved. The newly measured masses, with a precision of $\approx$ 1 keV/c$^2$, have been used to investigate the $N =$ 50 neutron shell closure confirming it to be robust. Empirical shell-gap and pairing energies determined with the new ground-state mass data are compared with the state-of-the-art \textit{ab initio} calculations with various chiral effective field theory Hamiltonians. The precise determination of the excitation energy of the $^{96m}$Ag isomer in particular serves as a benchmark for \textit{ab initio} predictions of nuclear properties beyond the ground state, specifically for odd-odd nuclei situated in proximity to the proton dripline below $^{100}$Sn. In addition, density functional theory (DFT) calculations and configuration-interaction shell-model (CISM) calculations are compared with the experimental results. All theoretical approaches face challenges to reproduce the trend of nuclear ground-state properties in the silver isotopic chain across the $N =$50 neutron shell and toward the proton drip-line. 
\end{abstract}

\maketitle


The nuclear landscape around the heaviest known doubly-magic self-conjugate nucleus $^{100}$Sn exhibits a rich variety of nuclear structure phenomena~\cite{FAESTERMANN2013}. Fundamental nuclear properties of nuclei in this region, such as their binding energies, are essential to probe the robustness of the $N$ = $Z$ = 50 shell closure and the evolution of single-particle energies~\cite{Hinke:2012vg}. These properties additionally support the investigation of the proton-neutron interaction on long-lived isomers~\cite{Boutachkov2011,Singh2011,DAVIES2017}, and the proximity of the proton drip line~\cite{Mazzocchi2007}.
Furthermore, the binding energies serve as important input for an accurate description of the rapid proton capture (rp) process occurring in extreme astrophysical environments~\cite{Schatz_2017,Zhou2023}. 
The robustness of the established magic numbers far from stability can be evaluated both with experimentally measured and theoretically estimated atomic masses~\cite{Erler:2012tu,Mougeot2021,Ali2023}. Masses of $^{95\textnormal{-}97}$Ag ($N$ = 48 - 50) provide information on the evolution of the $N$ = 50 shell gap near the proton dripline.
The question of whether the neutron magic number persists or diminishes remains open, and this phenomenon has been investigated in other regions~\cite{Wienholtz2013a,Michimasa18,Leis18}.
The atomic masses  near the $N$ = $Z$ line also provide information concerning the proton-neutron interaction and Wigner energy~\cite{Mardor21,Hamaker2021}. 
Computationally costly theoretical approaches to calculate the properties of neutron-deficient nuclei near the $N$ = $Z$ region have been intensively used to predict the binding energy, decay rates, excitation energies, quadrupole collectivity and the enhanced magicity~\cite{Caurier05,Hinke:2012vg,Lorusso2012,Togashi2018,Lubos2019,soma20,Park2020,HORNUNG2020,Hamaker2021,Mougeot2021,Sun2021,Zuker2021}. Accurate experimental mass data are highly desired to validate the predictions of different theoretical approaches. 


In this Letter, we present high-precision mass measurements of ground-state $^{95-97}$Ag nuclei as well as the isomeric state of $^{96}$Ag, using the JYFLTRAP double Penning trap mass spectrometer \cite{Eronen2012a} located at the Ion Guide Isotope Separator On-Line (IGISOL) facility~\cite{moore_igisol_2014}, within the Accelerator Laboratory of the University of Jyv{\"a}skyl{\"a}. The new mass values are used to quantify the robustness of the $N$ = 50 shell closure in the silver isotopic chain, and provide an opportunity to benchmark state-of-the-art \textit{ab initio} methods, density functional theory and shell model calculations close to the $N$ = $Z$ line.

To access nuclei in this region of the nuclear landscape, efficient and selective methods must be employed to produce exotic ion beams of the isotopes/isomers of interest, combined with sensitive detection techniques. In this work we have used a tailor-made approach, combining heavy-ion fusion-evaporation reactions with efficient laser resonance ionization in a hot-cavity catcher \cite{Reponen15} and sensitive ion detection via the phase-imaging ion-cyclotron resonance (PI-ICR) technique \cite{Eliseev2013,Nesterenko2018,Nesterenko2021}. This combination of production and detection methods was recently commissioned in a measurement of the mean-square charge radius of $^{96}$Ag, with a detection rate of about one ion every five minutes~\cite{reponen_evidence_2021}. 

\begin{figure*}[!htp]
\centering
   \includegraphics[width=1.99\columnwidth]
   {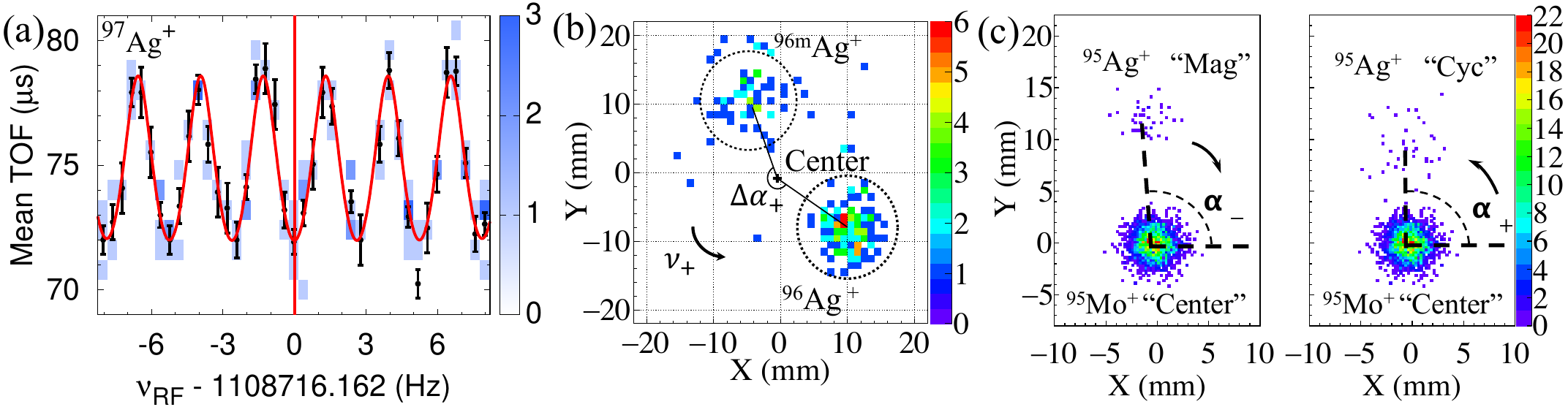}
   \caption{ 
(a) A Ramsey time-of-flight ion-cyclotron resonance (TOF-ICR) spectrum for $^{97}$Ag$^{+}$ acquired with a 25 ms (On) - 350 ms (Off) - 25 ms (On) excitation pattern. The black dots with uncertainties are the average TOF, and the solid line in red is the fit of the theoretical line shape. The vertical red line demonstrates the central frequency. 
(b) Well-resolved $^{96,96m}$Ag$^{+}$ ion radial motion (reduced cyclotron) projections on the 2-dimensional position-sensitive detector. (c) Ion spots (center, $^{95}$Mo$^{+}$, cyclotron phase and magnetron phase of $^{95}$Ag$^{+}$) after a typical PI-ICR excitation pattern with an accumulation time of 370 ms. The magnetron phase spot is displayed on the left side  and the cyclotron phase spot on the right. The angle difference between the two spots relative to the center spot is used to deduce the cyclotron frequency of the measured ion species. The number of ions in each pixel is illustrated by color bars.
}  
\label{fig:Ramesy-2-phases}
\end{figure*}
The $^{95-97}$Ag isotopes were produced using three reactions: $^{58}$Ni($^{40}$Ca, p2n)$^{95}$Ag, $^{60}$Ni($^{40}$Ca, p3n)$^{96,96m}$Ag and $^{92}$Mo($^{14}$N, $^{4}$He5n)$^{97}$Ag, and the hot-cavity catcher laser ion source was employed for efficient stopping and fast extraction. The reaction products recoiling out of the target were implanted into the hot graphite catcher from where they promptly diffused out into the catcher cavity as atoms. The silver atoms effused into a transfer tube and were selectively and efficiently ionized with a three-step resonant laser ionization scheme~\cite{reponen_evidence_2021}. After guidance through a radiofrequency sextupole ion guide~\cite{Karvonen2008}, the ions were accelerated to a potential of 30 kV into the mass separator. Reference ions of $^{95-97}$Mo$^+$ were separately generated either with an offline glow-discharge ion source \cite{Vilen2020b} or via sputtering from molybenum-made catcher components. 
An electrostatic bender was used to select ions either from the hot cavity or the offline ion source, prior to the first mass separation stage which is performed using a dipole magnet with a mass resolving power $\approx$ 500 
(in $\sigma$), 
sufficient for isobar separation. The selected ions were transported into a helium-gas filled radiofrequency-quadrupole cooler-buncher (RFQ-CB)~\cite{Nieminen2001} with a typical pressure of a few mbar. After cooling and bunching in the RFQ-CB, the ions were further directed to the JYFLTRAP double Penning trap,
consisting of two cylindrical Penning traps inside a 7-T superconducting solenoid. In the first (preparation) trap the buffer-gas cleaning method~\cite{Savard1991}  was applied, and an additional Ramsey cleaning cycle using dipolar excitation in the second (precision) trap~\cite{Eronen2008a} was used. 
This was to remove contaminant ions accompanying the $^{95-97}$Ag$^+$ ions allowing for measurements on minutely produced species and reducing ion-ion interaction related frequency shifts. In this work, the rate of $^{95}$Ag at JYFLTRAP was about one ion per 10 minutes. For the offline-produced reference ions, no additional purification was needed.

\begin{table*}[!htbp]
   \caption[]{ {Summary of the mass values obtained in this work.}
   Columns 1-8: isotope of interest, half-life, measurement method, ions of the reference nuclides used for the calibration, determined experimental frequency ratio $R$, the resulting mass excess (ME):  $M$ (atomic mass) – $A$ (atomic mass number) $\times$ $u$ (atomic mass unit) from this work and the literature (AME2020), and their difference ($\Delta$ME). Values marked with {\#} are extrapolated values from AME2020~\cite{Huang2021,Wang2021}.
  }
  \begin{ruledtabular}
   \begin{tabular*}{\textwidth}{cccccccc}
   %
 Isotope &Half-life (s) &Method &Reference &$R$ &ME (keV) this work &ME (keV) literature & $\Delta$ME (keV)\\
 \hline
 $^{95}$Ag&1.78(6)&PI-ICR&    $^{95}$Mo$^{+}$&1.000 316 374(16)   &-59743.3(14)&-59910{\#}(400{\#})&166{\#}(400{\#})\\
 $^{96}$Ag&4.45(3)&PI-ICR&    $^{96}$Mo$^{+}$&1.000 270 201 2(95)&-64656.69(95) &-64510(90)&-146(90)\\
 $^{96m}$Ag&6.9(5)&PI-ICR&   $^{96}$Mo$^{+}$&1.000 271 488(17)   &-64541.7(15) &&\\
 $^{97}$Ag&25.5(3)&TOF-ICR& $^{97}$Mo$^{+}$&1.000 184 004(16)  &-70935.3(14) &-70904(12)&-31(12)\\
   \end{tabular*}
   \label{table:mass}
   \end{ruledtabular}
\end{table*}

The masses of the ion of interest,  $^{95,96,96m,97}$Ag$^+$, were directly determined by measuring the ion’s cyclotron frequency $\nu_{c}$ in the precision trap~\cite{Eronen2012a,Nesterenko2021}. This frequency is given by the formula: $\nu_{c}=\frac{1}{2\pi}\frac{q}{m}B$, where $B$ represents the magnetic field strength, $q$ is the charge state and $m$ the mass of the ion.
$B$ is calibrated by evaluating the $\nu_{c}$ of corresponding isobaric reference ions, specifically $^{95,96,97}$Mo$^+$, both before and after each measurement of the ion of interest.

Two methods have been used to measure the $\nu_{c}$ of the ions in this work. The first is a time-of-flight ion-cyclotron resonance (TOF-ICR) technique~\cite{Koenig1995,Graeff1980}, which has been conventionally used to determine $\nu_{c}$ with either a single quadrupole excitation or the so-called Ramsey excitation. 
In this experiment, we used Ramsey-type time-separated oscillatory fields for the quadrupolar excitation~\cite{George2007a,George2007,Kretzschmar2007}. The quadrupolar excitation was applied as two 25-ms fringes separated by 350 ms for $^{97}$Ag$^{+}$ and  $^{97}$Mo$^{+}$. 
A typical Ramsey-type TOF-ICR spectrum obtained for $^{97}$Ag$^{+}$ is shown in Fig. \ref{fig:Ramesy-2-phases}(a).
The second method was the phase-imaging ion-cyclotron-resonance (PI-ICR) technique~\cite{Eliseev2013,Eliseev2014,Nesterenko2018,Nesterenko2021}, which relies on projecting ion motion within the Penning trap onto a position-sensitive ion detector (see ~\cite{Nesterenko2018,Nesterenko2021} for details). Measurement scheme number 2 described in~\cite{Eliseev2013} was applied to measure the cyclotron frequency $\nu_{c}$ of the ions. This scheme enables the direct measurements of the angle between the accumulated phases of cyclotron and magnetron motions ($\alpha_+$ and $\alpha_-$) with respect to the center spot:   $\alpha_c = \alpha_+-\alpha_-$. The cyclotron frequency can be deduced from: 
$\nu_{c}=({\alpha_{c}+2\pi n_{c}})/({2\pi{t_{acc}}}),$ 
where $n_{c}$ is the number of complete revolutions of the ions during the phase accumulation time $t_{acc}$. The chosen duration $t_{acc}$ ensured the separation of the cyclotron spot of the target ion from any potential isobaric, isomeric, or molecular contaminants. A clear separation between the ground and isomeric states of $^{96}$Ag is illustrated in Fig. \ref{fig:Ramesy-2-phases}(b). 
The accumulation time was adjusted to closely align with integer multiples of the $\nu_c$ period, thus ensuring that $\alpha_c$ remained small to reduce systematic shifts caused by image distortion. 
In addition, the delay of the cyclotron motion excitation was repeatedly scanned over one magnetron period and the final extraction delay was varied over one cyclotron period to account for any residual magnetron and cyclotron motions that could shift the different spots. 
An example of PI-ICR data illustrating the phase-imaging spots for the measurement of $^{95}$Ag$^{+}$ is shown in Fig. \ref{fig:Ramesy-2-phases}(c). 
The atomic mass of the ion of interest was derived from the measured cyclotron frequency ratio: $M_{ioi} = R(M_{ref} - m_e)   + m_e  + (R \cdot B_{ref} - B_{ioi})/c^2$, where $M_{ioi}$ and  $M_{ref}$ represent the masses of the ion of interest and reference ion, respectively. $R = {\nu_{c,{ref}}}/{\nu_{c,ioi}}$ denotes the cyclotron frequency ratio for singly charged ions. $m_{e}$ is the mass of the electron and $c$ the speed of light in vacuum. $B_{ref}$ and $B_{ioi}$ are the electron binding energies of the atoms of interest and reference atoms, which are a few eV~\cite{NIST_ASD} and can be neglected. Temporal fluctuations in the Penning trap magnetic field between consecutive reference measurements of a few minutes are considered in the analysis~\cite{Nesterenko2021}. Since the selected reference ion species have the same $A/q$ as the ion of interest, mass-dependent shifts effectively become inferior compared to the statistical uncertainty achieved in the measurements. The analysis of both methods revealed no count-rate-related frequency shifts, as the rates of the ions of interest were around 0.02, 0.01, 0.004, and 0.001 counts per bunch for $^{97}$Ag, $^{96}$Ag, $^{96m}$Ag, and $^{95}$Ag.
The final weighted mean frequency ratio $\overline{R}$ and the deduced  atomic masses are summarised and compared to the Atomic Mass Evaluation 2020 (AME2020)~\cite{Huang2021,Wang2021} values in Table~\ref{table:mass}.

The mass of $^{95}$Ag was measured for the first time and is 166(400)~keV less bound compared to the AME2020 extrapolation. The mass precision for $^{96}$Ag was improved by almost two orders of magnitude and it was found to be 146(90)~keV (1.6$\sigma$) more bound than in AME2020,  where an indirect measurement was used to evaluate its mass. The excitation energy of $^{96m}$Ag (2$^+$) was determined to be 115.0(17) keV. The new mass value for $^{97}$Ag is a factor of 9 more precise and deviates from AME2020~\cite{Huang2021,Wang2021} by 2.6$\sigma$. Our result is in good agreement with the reported value in ~\cite{Ali2023} but around 14 times more precise. 

\begin{figure*}[htb]
\begin{center}
\includegraphics[angle = 0,width=0.98\textwidth]{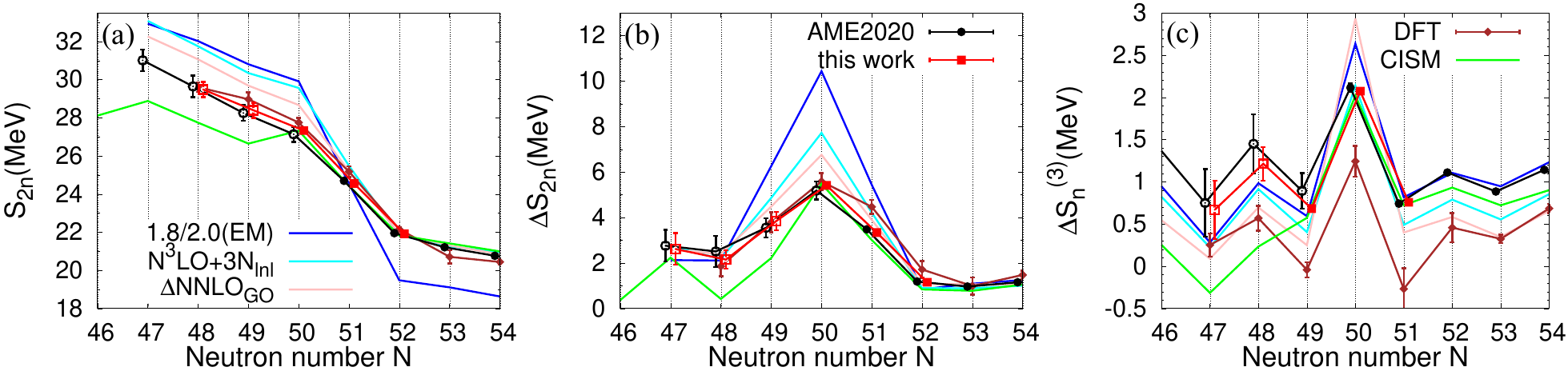}
\end{center}
\caption{ 
 (a) $S_{2n}$, (b) $\Delta S_{2n}$ and (c) $\Delta S_{n}^{(3)}$ of Ag isotopes as a function of neutron number. Our results (red squares) are compared to the evaluated values from AME2020 (black circles), and predictions from  DFT (brown), CISM (green), and \textit{ab initio} calculations with various Hamiltonians (cyan:  N$^3$LO+3N$_{\rm lnl}$; blue:  1.8/2.0(EM); pink: $\Delta$NNLO$_{\rm GO}$). If data from this work or experimental data from AME2020 include the extrapolated data, the data points are indicated with open squares or open circles. The error bars illustrate 1$\sigma$ uncertainty of the data points.
}
\label{fig:s2n}
\end{figure*}

As one moves along an isotopic chain, changes in the nuclear structure can be probed via irregularities in the generally smooth trends of two-neutron separation energies, $S_{2n}(N,Z) = \left[ME(Z,N-2) + 2ME_n - ME(Z,N)\right]c^2$, where $ME(Z,N)$ represents the mass excess of an isotope with $N$ neutrons and $Z$ protons and $ME_n$ is the neutron mass excess.
The changes in the slopes of the two-neutron separation energies are emphasized by looking at the derivative of $S_{2n}$, namely the two-neutron empirical shell gap: ${\Delta S_{2n}(Z,N)  =  S_{2n}(Z,N)  -  S_{2n}(Z,N+2)}$. $\Delta S_{2n}$ reflects the difference in the single-particle structure. Figure~\ref{fig:s2n}{(a) and (b)} present the experimental $S_{2n}$ and $\Delta S_{2n}$ values from this work, respectively, comparing to existing AME2020 values. With our high-precision mass measurement of $^{95}$Ag, we can determine the empirical neutron shell gap at $N = 50$ experimentally for the first time. This, combined with the updated $^{96,97}$Ag masses, enables characterization of the ground-state binding energy trend across the $N$ = 50 shell in silver. The new $S_{2n}$ values for $^{97}$Ag and $^{98}$Ag are 194.0 keV higher and 145.1 keV lower than the adopted values in AME2020, resulting in the same magnitude of change of the empirical neutron shell-gap energies. 
The $S_{2n}$ value from this work for $^{99}$Ag is in good agreement with the value from AME2020. 
The determined empirical shell gap energy at $N$ = 50, $\Delta S_{2n}$($^{97}$Ag) = 5414.7(67) keV, is more than 200 keV larger than that predicted by AME2020, and a distinct peak appears at $N$ = 50 as illustrated in Fig.~\ref{fig:s2n}{(b)}, giving evidence for the existence of a robust shell closure. 

Another relevant feature related to nuclear structure is pairing. One of the most robust signatures of pairing in nuclei is the odd–even staggering (OES), the so-called pairing gap, which can be derived from neutron separation energies, ${ S_{n}(N,Z) = \left[ME(Z,N-1) + ME_n - ME(Z,N)\right]c^2}$, as ${\Delta S_{n}^{(3)}  =  (-1)^N(S_{n}(Z,N)  -  S_{n}(Z,N+1))/2}$ (see Fig.~\ref{fig:s2n}{(c)}).

We compare our new experimental data with state-of-the-art nuclear models (see Fig.~\ref{fig:s2n}), namely \textit{ab initio} valence-space in-medium similarity renormalization group (VS-IMSRG) calculations \cite{PhysRevLett.118.032502,Stro19ARNPS,Hu22Pb208} using the 1.8/2.0(EM) \cite{PhysRevC.83.031301,Simo17SatFinNuc}, N$^3$LO+3N$_{\rm lnl}$ \cite{Soma20LNL}, and $\Delta$NNLO$_{\rm GO}$ \cite{PhysRevC.102.054301} interactions, the configuration-interaction shell model (CISM) \cite{Liu2023} and the density functional theory (DFT) approach with the UNEDF0~\cite{PhysRevC.82.024313} Skyrme energy density functional.


The \textit{ab initio} VS-IMSRG~\cite{Stro19ARNPS} model allows us to test nuclear forces in fully open-shell systems. The calculations start from several two- (NN) and three-nucleon (3N) interactions based on chiral effective field theory \cite{Epel09RMP,Mach11PR}. The 1.8/2.0(EM) and N$^3$LO+3N$_{\rm lnl}$ interactions are fit to only $A \le 4$ data, but generally reproduce well ground-state energies up to the medium-mass region~\cite{Morr18Tin,Holt21Drip,Soma20LNL}. In contrast, $\Delta$NNLO$_{\rm GO}$ includes explicit $\rm{\Delta}$-isobar degrees of freedom and optimizes NN and 3N forces simultaneously at the N$^2$LO level to reproduce few-body data as well as saturation of infinite matter. Working in an initial harmonic-oscillator basis of 15 major shells at oscillator frequency $\hbar\omega$ = 16 MeV, 
we impose an additional cut on 3N matrix elements storage $E_{\rm 3max}$ = 24 \cite{PhysRevC.105.014302}. We first transform to the Hartree-Fock basis, then use the VS-IMSRG \cite{PhysRevLett.118.032502,Stro19ARNPS} to construct approximate unitary transformations to decouple a core and multi-shell valence-space Hamiltonian from the full $A$-body Hamiltonian~\cite{Miya20lMS}. 
We capture effects of 3N forces between valence nucleons via the ensemble normal ordering procedure \cite{PhysRevLett.118.032502}, such that a specific valence-space Hamiltonian is constructed for each nucleus to be studied. In this work we employ the $^{80}$Zr core with a proton $g_{9/2}$ and neutron $sdg$ valence space in order to cross $N = $ 50 for the silver isotopes. The final exact diagonalization was performed using the \textsc{KShell} shell-model code \cite{SHIMIZU2019372}. The theoretical uncertainty in the present calculations mainly arises from differences in the three chiral interactions, in particular differences in the partial wave $^1S_0$ of the nucleon-nucleon force which plays a central role in setting up pairing correlations in nuclei \cite{ekstrom2023chiral}.

%


The VS-IMSRG predictions of $S_{2n}$ and $\Delta S_{2n}$ follow the trend of the experimental values rather well but there are offsets of a few MeV, in particular for the 1.8/2.0(EM) interaction, and also for the N$^3$LO+3N$_{\rm lnl}$ and $\Delta$NNLO$_{\rm GO}$ interactions in the regions below $N=50$ (see Fig.~\ref{fig:s2n}(a) and (b)). The observed offsets are not unusual as discrepancies of a few MeV in the $S_{2n}$ values have also been observed in other medium-mass regions \cite{PhysRevC.101.014318}, potentially arising from the uncertainty associated with nuclear forces. The calculations using the $\Delta$NNLO$_{\rm GO}$ interaction agree best with the experimental $S_{2n}$ data and are closest to the experimental $\Delta S_{2n}$ at $N=50$. All \textit{ab initio} calculations overpredict the two-neutron shell gap at $N=50$. 

While the $\Delta$NNLO$_{\rm GO}$ interaction works reasonably well for the two-neutron shell-gap energies (see Fig.~\ref{fig:s2n}.(b)), it does not reproduce the magnitude of odd-even effects so well (see Fig.~\ref{fig:s2n}.(c)). It clearly overestimates the pairing gap at $N$ = 50 and underestimates pairing effects in all other regions. The N$^3$LO+3N$_{\rm lnl}$ interaction agrees very well with the experimental pairing gap at $N$ = 50 but, similar to $\Delta$NNLO$_{\rm GO}$, underestimates it in the other regions. The 1.8/2.0(EM) interaction shows an excellent agreement with the experimental data above $N$ = 50 but for $N\leq50$ it behaves similar to $\Delta$NNLO$_{\rm GO}$. 
In conclusion, the IMSRG(2) approximation used in this work, where all operators are truncated at the two-body level, systematically overestimates the strength of the shell closure feature \cite{Taniuchi2019}. 




The CISM \cite{Liu2023} calculations
were performed using the Hamiltonian in combination with jj45pna \cite{HJORTHJENSEN1995125} and VMU \cite{Otsuka2010} + LS \cite{BERTSCH1977399} interactions
in the model space of proton $f_{5/2}pg_{9/2}$ orbitals and neutron $f_{5/2}pgdsh_{11/2}$ orbitals. 
As shown in green in Fig.~\ref{fig:s2n}(a)-(c), the CISM calculations agrees well with the experimental data for $N=50-54$. The $N$ = 50 shell gap is well reproduced but for $N\le$50, the trends are different from the experimental data, including our new values and the extrapolated values from AME2020 \cite{Wang2021}. While our work and the \textit{ab initio} calculations indicate a smooth linear decrease in the $S_{2n}$ values for $N=47-50$, suggesting the absence of a sub-shell closure below $N$ = 50, the CISM indicates a change in nuclear structure at $N=47$. To further explore the potential presence of an augmented deformed shell closure and possibly elevated Wigner energy at $N=Z=47$, as exemplified in \cite{Hamaker2021}, the atomic masses of silver isotopes extending beyond $N=47$ are desired.

The DFT calculations were performed in the same manner as in Ref.~\cite{reponen_evidence_2021}.  
Namely, we used the computer code HFBTHO~\cite{Stoitsov2013}, which solves Hartree-Fock-Bogoliubov equations in an axially symmetric harmonic oscillator basis, and the UNEDF0~\cite{PhysRevC.82.024313} Skyrme EDF. The odd-particle nuclei were computed with the quasiparticle blocking procedure, by using the equal filling approximation~\cite{Perez-Martin2008}. To assess the predictive power of the model, we have computed propagated statistical errors by utilizing the covariance matrix of the EDF model parameters~\cite{PhysRevC.87.034324,Haverinen_2017}. 
The calculations (brown triangles in Fig.~\ref{fig:s2n}) agree with the experimental $S_{2n}$ values rather well and produce an excellent agreement with the two-neutron shell-gap energy at $N$ = 50. 
However, the DFT model systematically underestimates the pairing gap (see Fig.~\ref{fig:s2n}.(c)), which may stem from too weak pairing correlations. 
The magnitude of the $\Delta S_{n}^{(3)}$ staggering is, however, well produced by the DFT model.

\begin{figure}[!htp]
\centering
   \includegraphics[width=0.99\columnwidth]{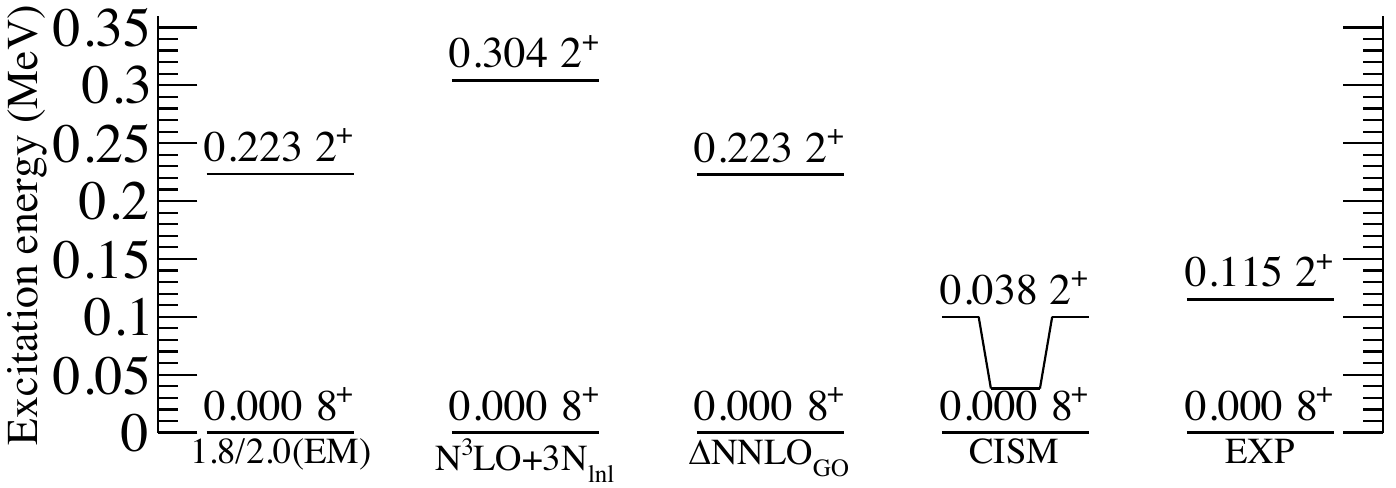}
   \caption{ 
Partial energy level scheme of $^{96}$Ag. The experimental data (EXP) is compared to calculated excited states using the 1.8/2.0(EM), N$^3$LO+3N$_{\text{lnl}}$ and $\Delta$NNLO$_{\text{GO}}$ interactions, as well as CISM.
}  
\label{fig:energy-level-96Ag}
\end{figure}
In this work, we also report the first successful identification 
and excitation energy measurement of the $\beta$-decaying isomer in $^{96}$Ag. 
The ground and isomeric states ($8^{+}$ and $2^{+}$, as experimentally assigned in~\cite{Park2019}) were distinguished through mass differences employing the PI-ICR technique. The $2^{+}$ isomeric state to the $8^{+}$ ground state ratio was determined to be 0.366(25), which  is consistent with the observation reported in~\cite{BATIST2003}.
In the used fusion-evaporation reaction, the cross-section for generating the $8^{+}$ state is considerably larger than that for the lower-spin $2^{+}$ isomeric state, aligning with the observed rate (Fig. \ref{fig:Ramesy-2-phases}(b))~\cite{Weber2008b,Bass80,BATIST2003}. 
We have performed \textit{ab initio} and CISM calculations, and compare their predictions to the excitation energy from this work in Fig.~\ref{fig:energy-level-96Ag}. CISM predicts the occurrence of $8^{+}$ and $2^{+}$ states through the coupling of three proton $\pi g_{9/2}$ holes and one $\nu g_{9/2}$ neutron hole, with a probability of 90\%. 
All models predict an $8^+$ ground state and a $2^+$ isomer, consistent with experimental findings. Notably, the theoretical assessments presented in~\cite{Becerril2011,Park2019} indicate inverted spins for these two states. The CISM excitation energy for the isomer is lower than the experimental value, whereas the \textit{ab initio} calculations somewhat overestimate it. The 1.8/2.0(EM) and $\Delta$NNLO$_{\rm GO}$ calculations agree with the experiment to approximately 100 keV, while N$^3$LO+3N$_{\rm lnl}$  predicts a somewhat higher energy, 304 keV. These minor discrepancies fall well within expected theoretical uncertainties stemming from interaction differences, many-body and model-space truncations, and the normal-ordering approximation for 3N forces. 

The odd-odd nucleus $^{96}$Ag, with three $\pi g_{9/2}$ proton holes and one $\nu g_{9/2}$ neutron hole below $N=50$, 
is particularly important for understanding the evolution along the $N$ = 49 isotones toward the single proton and neutron hole nucleus $^{98}$In. 
Odd-odd nuclei serve as a complementary and rigorous testing ground for \textit{ab initio} approaches which are predominantly focused on even-even and odd-A nuclei, especially in the context of drip-line nuclei. Presently, the application of \textit{ab initio} calculations to odd-odd nuclei, especially beyond the ground state, is very limited, primarily due to the complex proton-neutron interaction involving unpaired protons and neutrons. The precise measurement of the isomeric excitation energy of $^{96m}$Ag is noteworthy, as it matches the results of \textit{ab initio} calculations conducted as benchmark tests on odd-odd nuclei near the proton dripline in the vicinity of $^{100}$Sn.
Similarly, level ordering in odd-odd $N=49$ $^{94}$Rh was recently verified through excitation energy measurements of its long-lived isomeric state in combination with theoretical calculations~\cite{Ali2023}, leading to the determination of an $8^+$ ground state in $^{94}$Rh.
Lastly, we note that $^{96m}$Ag may function as an astrophysical nuclear isomer, referred to as an "astromer", maintaining its metastable nature in relevant astrophysical conditions~\cite{Misch2021,Misch24}. This marks the first precise measurement of the excitation energy of the isomer of $^{96}$Ag, enabling the ground state and isomer of $^{96}$Ag to be treated as separate species in astrophysical modeling.

We have reported direct mass measurements of exotic silver isotopes $^{95-97}$Ag close to the proton drip line. The two-neutron empirical shell gap for $N = 50$ has been determined as 5414.7(67) keV, signifying the most elevated shell gap energy within the experimentally known range from $Z = 42$ up to $Z = 50$.
Our new experimental data have enabled the benchmarking of calculations from \textit{ab initio} theory, configuration-interaction shell-model and density functional theory for the proton-rich silver isotopes. 
These high-precision measurements provide crucial information for the development of nuclear forces to extend these methods to higher accuracy. 
Our work highlights the scientific capabilities of the new experimental method employed here for the first time. The coupling of the PI-ICR Penning-trap mass spectrometer technique with the hot cavity catcher laser ion source provides remarkably high sensitivity that enables the direct mass measurements of exotic isotopes with a production rate of $\approx$1 count per 10 minutes. 
Future precision mass data for $^{94}$Ag at the $N$ = $Z$ line would provide important constraints to guide the ongoing development of \textit{ab initio} and other theoretical approaches for nuclear structure and help to understand the increased binding predicted by the CISM for $^{94}$Ag.

We acknowledge the staff of the Accelerator Laboratory of University of Jyv\"askyl\"a (JYFL-ACCLAB) for providing stable online beam. We express gratitude for the productive results stemming from conversations with X. F. Yang. We thank the support by the Academy of Finland under the Finnish Centre of Excellence Programme 2012-2017 (Nuclear and Accelerator Based Physics Research at JYFL) and projects No. 306980, No. 312544, No. 275389, No. 284516, No. 295207, No. 314733, No. 315179, No. 327629, No. 320062, No. 339243, No. 354589, No. 345869 and No. 354968. The support by the European Union’s Horizon 2020 research and innovation program under grant No. 771036 (ERC CoG MAIDEN) and No. 861198–LISA–H2020-MSCA-ITN-2019 is acknowledged.  
The VS-IMSRG calculations were supported by the U.S. Department of Energy under No.~DE-FG02-96ER40963 and SciDAC-5 (NUCLEI collaboration), and the Natural Sciences and Engineering Research Council of Canada under grants SAPIN-2018-00027 and RGPAS-2018-522453, as well as the Arthur B. McDonald Canadian Astroparticle Physics Research Institute and were performed with an allocation of computing resources on Cedar at WestGrid and The Digital Research Alliance of Canada. Support provided by the National Natural Science Foundation of China (No. 11775277) is appreciated. We acknowledge the CSC-IT Center for Science Ltd., Finland, for the allocation of computational resources. 
This work was supported by the German Federal Ministry for Education and Research (BMBF) under contracts no. 05P21RGFN1, by the German Research Foundation (DFG) under contract No. SCHE 1969/2-1, by HGS-HIRe, and by Justus-Liebig-Universität Gießen and GSI under the JLU-GSI strategic Helmholtz partnership agreement. The paper was supported by the DAAD Grant No. 57610603.
Z.~Ge and M.~Reponen contributed equally to this work. 

\putbib
\end{bibunit}

\end{document}